\documentclass[aps,prl,twocolumn,showpacs,floatfix,a4paper]{revtex4}
\usepackage{graphicx}
\usepackage[intlimits]{amsmath}

\usepackage{amsfonts}
\usepackage{amssymb}
\usepackage{verbatim}
\usepackage{amsmath}
\usepackage{mathrsfs}

\usepackage{color}

\definecolor{pink}{rgb}{1,0.39,0.8}
\definecolor{grey}{rgb}{0.8,0.8,0.8}

\begin{document}

\title{Compressibility enhancement in an almost staggered interacting 
Harper model}
\author{Bat-el Friedman and Richard Berkovits}
\affiliation{Department of Physics, Jack and Pearl Resnick Institute, Bar-Ilan
University, Ramat-Gan 52900, Israel}

\begin{abstract}
We discuss the compressibility in the almost staggered fermionic 
Harper model with repulsive interactions in the vicinity of half-filling. 
It has been shown by Kraus et al. \cite{kraus14} that 
for spinless electrons and nearest neighbors electron-electron interactions 
the compressibility in the central band is enhanced by repulsive interactions. 
Here we would like to investigate the sensitivity of this conclusion to
the spin degree of freedom and longer range interactions. 
We use the Hartree-Fock (HF) approximation, as well as density matrix 
renormalization group (DMRG) calculation to evaluate the compressibility.  
In the almost staggered Harper model, the central energy band is essentially 
flat and separated from the other bands by a large gap and therefore, the 
HF approximation is rather accurate. 
In both cases the compressibility of the system is enhanced compare to 
the non-interacting case, although the enhancement is weaker due to the 
inclusion of Hubbard and longer ranged interactions. We also show that
the entanglement entropy is suppressed when the compressibility of 
the system is enhanced.
\end{abstract}


\maketitle

\section{Introduction}

The interplay between
electron-electron (e-e) interactions and quasi-disorder 
has drawn much excitement since the discovery of quasi-crystals
\cite{shechtman84,levine84}.  
Much of the work has focused on a specific model of a one-dimensional 
(1D) quasi-crystal, namely
the Harper (or Aubry-Andr\'{e}) model \cite{Harper:1955,AA}.
One of the main attractions of this model is that contrary to conventional
1D disordered systems which are localized for any amount of 
disorder \cite{lee85}, 
the Harper model exhibits a metal-insulator transition
as function of the quasi-disordered potential strength, 
even in the absence of interactions 
\cite{AA,hiramoto92,Svetlana,Roati08,Yoav09,Chabe08,Modugno10}.
The influence of e-e interactions on the metal-insulator transition
of the Harper model was studied in several publications 
\cite{vidal99,schuster02,iyer13}.
Interest in the Harper model has lately peaked after it has been
shown that for an irrational modulation,
the Harper model may be a 1D topologically nontrivial system, and have
topological boundary states
~\cite{kraus12,kraus12a,madsen13,verbin13,xu13,ganeshan13,grudst13,xu13a,satija13}.
This property,
coupled with the fact that the Harper model may be realized in the context
of cold atoms and molecules 
\cite{Bloch:2013,Ketterle:2013}
added to the excitement surrounding the Harper model.

Recently, an additional aspect of the model has been investigated, namely the 
inverse compressibility, which measures the change in the chemical potential 
when an electron is added to the system. In the context of disordered quantum
dots this has become a very popular measurement to extract information on
the role of e-e interactions in these systems 
\cite{sivan96,berkovits98,alhassid00}. 
For a finite  system of $N$ particles, $\Delta_2(N)$, is defined as the 
change in the chemical potential due to the insertion of the $N^{\rm th}$ 
particle
i.e., $\Delta_2(N)=\mu(N)-\mu(N-1)$, where $\mu(N)$ is the chemical potential
for $N$ particles. Since $\mu(N)=\mathcal{E}(N)-\mathcal{E}(N-1)$ 
(where $\mathcal{E}(N)$
is the system's many-body ground-state energy with $N$ particles),  
$\Delta_2(N)$  is given by:
\begin{align} \label{Eq:Delta2}
\Delta_2(N) = \mathcal{E}(N) - 2\mathcal{E}(N-1) + \mathcal{E}(N-2) \,,
\end{align}
For non-interacting systems at zero temperature,
\begin{equation}
\Delta_2(N) = E_N - E_{N-1} = \Delta(N), 
 \label{Eq:D2}
\end{equation}
where $E_N$ is the $N^{\rm th}$ single-particle eigenenergy 
and $\Delta(N)$ is the 
single-particle level spacing.

How do the e-e interactions affect the inverse compressibility? 
The conventional wisdom
leads to the constant interaction (CI) model \cite{alhassid00,kurland00},
which essentially assumes that the interactions between the electrons 
are well described by mean-field. This leads to the conclusion that the
effect of interactions on the inverse participation given by
$\Delta_2(N) = \Delta(N) + e^2/C$, where $C$ is the total classical 
capacitance.
Thus, the e-e interactions increase the inverse compressibility compared to its
non-interacting value. This description fits well the experimental 
measurements in quantum dots~\cite{alhassid00}.  

However, the CI mean-field 
description doesn't hold at certain conditions.
It has been shown \cite{usuki,furukawa,assaad} that close to the 
Mott metal-insulator transition occurring at half-filling of a clean 
the Hubbard model, the inverse compressibility 
may decreases with the Hubbard 
interactions strength. Recently, it has been shown 
\cite{kraus14} that for the almost staggered Harper model of 
spinless electrons with nearest-neighbors e-e interactions, close to 
half-filling, the system becomes more compressible 
as the interactions are increased, although no metal-insulator transition
occurs there.
This counter intuitive behavior stems from the properties of the 
electronic bands and density 
for the almost staggered Harper model. Under these conditions 
the non-interacting 
Harper model has an almost flat narrow band around zero energy, 
separated from the other
bands by large gaps. The density of the narrow band around half-filling 
is anti-correlated 
with the on-site potential, whereas the density of the lower 
occupied bands follows the potential.
Therefore, once e-e interaction is introduced, 
the electrons in the lower occupied bands squeeze
out the states in the narrow central band, resulting in a
narrower central band. This flattening of the central band 
due to the interaction with
the lower band electronic density results in an increase 
of the compressibility.

In this paper we address the question whether this increase 
of the compressibility
is the result of the particular model studied in Ref. \cite{kraus14}.
Specifically, we shall see what happens to the compressibility 
when the spin degree
of freedom is taken into account, or equivalently when considering 
a spinless two legged ladder. Another case which we explore 
is when next nearest neighbors interactions are included.
To study the compressibility we mainly rely on the HF approximation,
which has been shown to be extremely accurate for this model \cite{kraus14} due
to the large gap between the flat central band and the lower band and 
to the localized nature
of the states in the narrow band. We will also compare some of these results
to density matrix renormalization group (DMRG) numerical calculations,
which for these 1D systems are essentially exact \cite{white92,dmrg},
and describe very well the dependence of the ground state energy on the
number of particles \cite{berkovits05}. Using DMRG we also show that
the enhancement of the compressiblity is accompanied by the suppression
of the entanglement entropy.

\section{Hubbard interaction}
\label{s1}

 In this section we discuss the influence of the spin degree of freedom 
on the compressibility in the staggered Harper model close to half-filling.
The clearest difference between spin-polarized (spinless) and non-polarized
electron is the fact that for non-polarized (spinfull) electrons
there are Hubbard  interactions.
The on-site potential is spatially modulated with a frequency of almost two 
lattice-sites period (i.e., staggered), corresponding to fast modulation with 
a slow envelope. The interaction terms are repulsive and short ranged 
(on-site and nearest-neighbors (n.n.)-interactions). We assume that in
the limit of weak Hubbard interactions no spin polarization occurs, i.e., 
the total $S_z=0$ for even filling and $S_z=\pm 1/2$ for odd filling.
We show that the compressibility of the system decrease 
when the Hubbard interactions are
increased by analyzing the central (flat) energy band close to half-filling.
Due to the Kramers degeneracy, as long as there is no spin flip 
(tunneling between the ladders' legs), the single-particle 
solution is just a duplication
of the spinless solution presented in Ref. \cite{kraus14}. 
Thus, it contains two copies 
of superlattice states that reside at the valleys of the 
potential envelope. 
Since the electrons are localized in the potential valleys, 
adding an additional electron 
to a valley will increase the energy due to the Hubbard interaction. 
In order to reduce the effect of the Hubbard interaction the 
electronic density must rearrange itself. 
As a result, the capacitance of the system goes down. 

In order to demonstrate that behavior we need to explicitly solve
the tight-binding Harper model for fermions with spin and with Hubbard 
and n.n. 
repulsive interactions given by:
\begin{equation}
\label{hamiltonian}
\begin{aligned}
H&=\sum_{s\neq s'=\uparrow,\downarrow}\sum_{j=1}^L \big[t(c^{\dagger}_{j,s}c_{j+1,s}+h.c.)+t'c^{\dagger}_{j,s}c_{j,s'}\\
&+\lambda \cos(2\pi bj+\phi)n_{j,s}+Un_{j,s}n_{j+1,s}+U'n_{j,s}n_{j,s'})\big].
\end{aligned}
\end{equation}
where $c_{j,s}$ is the single particle annihilation operator at 
site $j$ with spin $s$ and 
$n_{j,s}=c^{\dagger}_{j,s}c_{j,s}$ is the number operator. 
$t,t' \in \mathbb{R}$ are the 
site hopping and spin flipping amplitudes, respectively. 
$\lambda>0$ controls the on-site 
potential amplitude. The potential is a cosine modulated in space 
with frequency $b$ and 
a phase factor $\phi$ . $U>0$ and $U'>0$ are the strengthens of the 
repulsive n.n. and Hubbard interactions, respectively. 
We discuss the region $\lambda<2t$, which is 
the metallic regime \cite{AA}. We further assume that 
$b \mod 1= 1/2+\epsilon$, 
$\epsilon\ll 1/2$ corresponding to an almost staggered case.  
$\epsilon\in \mathbb{R}$ is non-rational so
that the system is disordered.

\begin{figure}[ht!]
\centering
\includegraphics[width=90mm]{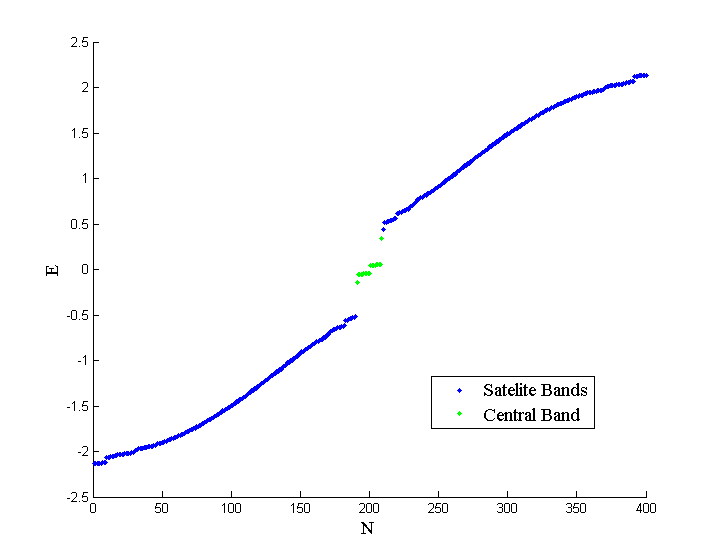}
\caption{Energy bands of the free Hamiltonian (U=U'=0). 
The central band is splitted due to the gap created by the spin 
flipping amplitude $t'$. The parameters used through the figures are 
$t=1;
t'=0.05;
\lambda=0.7 ;
\phi=0.7\pi;
b=\sqrt{30};\epsilon = -0.0228;L=200.$ The isolated points 
correspond to protected edge (topological) states in the Harper model 
and are not discussed in this paper.}
\label{spinebands}
\end{figure}

Let us first discuss the non-interacting Hamiltonian, i.e., 
set $U,U'=0$ in Eq. (\ref{hamiltonian}).  
%
A numerical solution in this case reveals the existence of
an almost flat 
central energy 
band (see Fig. \ref{spinebands}), 
splitted due to the spin flip matrix element to a lower and higher central
band. 
We are mostly interested in 
the central band energy spectrum, and since these energy states 
which are close to zero minimize both kinetic 
and potential energy, we conclude that the most important contribution 
comes from states 
localized in the potential valleys, i.e. states localized around the 
position $l_z$ corresponding
to $2\pi\epsilon l_z+\phi= (\mathbb{Z}+\frac{1}{2})\pi$ \cite{kraus14}. 
In the valley, we can approximate 
$\cos(2\pi\epsilon j+\phi)\approx 2\pi|\epsilon|(j-l_z)s_z$, and 
$s_z=-sign(\sin(2\pi\epsilon l_z+\phi))=\pm 1$.  
The effective Hamiltonian describing the central band is
\begin{equation}
\begin{aligned}
&H^{val}=\sum_{s\neq s'=\uparrow,\downarrow}\sum_{j=1}^L\big[t(c^{\dagger}_{j,s}c_{j+1,s}+h.c.)+\\
&t' c^{\dagger} _{j,s}c_{j,s'}+2\pi\epsilon\lambda s_z(-1)^j(j-l_z)c^{\dagger}_{j,s}c_{j,s}\big]\\
&=\sum_{s,s'=\uparrow,\downarrow} \frac{L}{2\pi}\int^{\pi}_0 \Psi^{\dagger}_{k,s}\big[(2t\cos(k)\sigma_x+\\
&2\pi|\epsilon|s_z\lambda(\hat p_k-l_z)\sigma_z)\delta^s_{s'}+t'(1-\delta^s_{s'})]\Psi_{k,s'}\\
\end{aligned}
\end{equation}
where \[\psi_{k,s}= \left( \begin{array}{ccc}
c_{ek,s} \\
c_{ok,s}\\
\end{array} \right),\] 
is the sub-lattice pseudo-spinor that splits the lattice 
into even and odd sites, 
according to $c_{ek,s}=\frac{2}{L}\sum_{j=1}^{L/2}e^{ik2j}c_{2j,s}$ 
and $c_{ok,s}=\frac{2}{L}\sum_{j=1}^{L/2}e^{ik(2j-1)}c_{2j-1,s}$. 
$\hat p_k\equiv i\partial_k$ and $\sigma_x,\sigma_z$ are the $2 \times 2$
Pauli matrices.

Diagonalizing the spin degrees of freedom 
(which are independent of k-space), we get 
\[\psi_{k,1}= \frac{1}{\sqrt{2}}\left( \begin{array}{ccc}
c_{ek,\uparrow}+c_{ek,\downarrow}\\
c_{ok,\uparrow}+c_{ok,\downarrow}\\
\end{array} \right)\]
\[\psi_{k,2}= \frac{1}{\sqrt{2}}\left( \begin{array}{ccc}
c_{ek,\uparrow}-c_{ek,\downarrow}\\
c_{ok,\uparrow}-c_{ok,\downarrow}\\
\end{array} \right).\]

This representation allows us to write the 
Hamiltonian as a sum of two distinct subspaces, 
each relates to a different spin eigenstate. 
The subspaces depend only on the momentum $k$, 
and therefore can be solved using the same methods 
used for spinless fermions \cite{kraus14}.
Thus, the eigenenergies for the Hamiltonian of the potential valleys are
$E_1^{val}=\pm \sqrt{8n}\dfrac{t}{\xi}+t'$, and 
$E_2^{val}=\pm \sqrt{8m}\dfrac{t}{\xi}-t'$, where $m,n \in \{0,1,2...\}$, 
and $\xi^2=\dfrac{t}{\pi\lambda|\epsilon|}$. 
$ E_1^{val}, E_2^{val}$ correspond to the spin states 
$1,2$ respectively. 
The central zero-energy band splits due to the spin flip, resulting 
in an energy splitting between the two bands equal to $2t'$.

The eigenfunctions for the states belonging to the splitted central band are:
\begin{equation}
|l_{z,i}>\approx (\pi\xi^2)^{-\frac{1}{4}}\sum_{j=1}^L (s_z)^j\bold{S}_j 
e^{-\frac{(j-l_z)^2}{2\xi^2}}|j,i>,
\end{equation}
where $|j,i>=\frac{1}{\sqrt{2}}(c_{j,\uparrow}^{\dagger} \pm 
c_{j,\downarrow}^{\dagger})|\emptyset> $
, where $|\emptyset>$ is the vacuum state.
These wavefunctions are Gaussians of width $\xi$ around $l_z$.
In the limit of small $t'$ our assumptions hold and this result 
is a good approximation of the real ground state.

These states form a basis for the central band, defined by $m,n=0$, since
$<l_{z,i}|l_{z\pm1,i}>=0, <l_{z,1}|l_{z,2}>=0$, and 
$|<l_{z,i}|l_{z',i}>|\le e^{-\frac{(l_z-l_{z'})^2}{2\xi^2}}\ll 1$.

Let us now consider the contribution of the overlap between 
the localized states in
the central band. The Gaussian decay of the localized states implies that 
the Hamiltonian matrix elements, $<l_{z,i}|H|l_{z',j}>$,
are not negligible only
between nearest neighbors states $|z-z'| = 1$.
Thus, the central band states follow an effective Hamiltonian:
\begin{equation}
\begin{aligned}
&H^{central}=-\bar t \sum^{L_z}_{z=1}\sum_{i=1,2}(-1)^zc^{\dagger}_{l_z,i}
c_{l_{z+1},i}\\
&+h.c.+t'c^{\dagger}_{l_z,i}c_{l_{z},i}.
\end{aligned}
\end{equation}
Diagonalizing this Hamiltonian yields the eigenstates
\begin{equation}
|k,i>=L_z^{-1/2}\sum^{L_z}_{z=1}S_ze^{ikz}|l_{z,i}>
\end{equation}
with eigenvalues $E^{central}(k)=-2\bar t \cos(k)\pm t'$.


Now, let us focus on the case where the Hubbard interactions 
in the Hamiltonian Eq. (\ref{hamiltonian}) are turned on ($U'\neq 0$),
but no longer range interactions are yet considered ($U=0$).
 %
For $U'\rightarrow\infty$ the model can be solved analytically.
In that limit only the interaction term is important. 
The eigenenergies are therefore $E=0$ and $E=U'$. The latter case occurs 
when two particles 
with opposite spins occupy the same site. This will cost infinite 
energy and therefore 
such states are decoupled from the theory. The remaining states 
contain a single particle per site.

Next, we consider the case where $U'$ is much bigger than the other 
energy scales in the theory, 
i.e. $U' \gg t,t',\lambda$. 
Using perturbation theory with $t$ as the perturbation parameter on 
the Hubbard model reveals that ferromagnetism is the lowest energy state. 
Adding $t'$ to the theory 
will not change the ground state, since the correction in $t'$ 
will be of at least third order 
in perturbation theory. 

As is discussed in Ref. \onlinecite{kraus14}, because the central 
band is essentially
protected by the large gaps to the other bands, the HF approximation results 
are very accurate.
Therefore, we approximate the Hubbard interaction using the HF method 
for interaction strength values smaller than 
these gaps $U'\ll\sqrt{8}\dfrac{t}{\xi}$.
\begin{equation}
\begin{aligned}
\sum_j n_{j,\uparrow}n_{j,\downarrow}\approx & 
\sum_j [<n_{j,\uparrow}>n_{j,\downarrow}+n_{j,\uparrow}<n_{j,\downarrow}>\\
&-<n_{j,\uparrow}><n_{j,\downarrow}>].
\end{aligned}
\end{equation}
Rewriting the Hamiltonian in Eq. (\ref{hamiltonian}) with $U=0$, 
and ignoring the constant 
term which is simply a shift in the energy, results in
\begin{equation}
\begin{aligned}
&H=\sum_{s\neq s'=\uparrow,\downarrow}\sum_{j=1}^L \big[t(c^{\dagger}_{j,s}c_{j+1,s}+h.c.)+t'c^{\dagger}_{j,s}c_{j,s'}+\\
&\big(\lambda \cos(2\pi bj+\phi)+U'<n_{j,s'}>\big)n_{j,s}\big].
\end{aligned}
\end{equation}
We find that the averaged electronic density between the 
valleys of potential is 
$<n_{j,s}>\approx\frac{1}{4}-\frac{1}{2}(-1)^j\bar{n}
(\frac{\lambda}{2t})\cos(2\pi\epsilon j+\phi))$ , 
with $\bar{n}(x)=\dfrac{x}{\pi \sqrt{1+x^2}}K(\frac{1}{1+x^2})$, 
and $K$ is the complete elliptical integral of the first kind.
Hence 
\begin{equation}
\begin{aligned}
\label{hfhamiltonian}
&H^{HF}=\sum_{s\neq s'=\uparrow,\downarrow}\sum_{j=1}^L \big[t(c^{\dagger}_{j,s}c_{j+1,s}+h.c.)\\
&+t'c^{\dagger}_{j,s}c_{j,s'}+\big(\lambda_{eff} \cos(2\pi bj+\phi)+\frac{1}{4}U'\big)n_{j,s}\big],
\end{aligned}
\end{equation}
where $\lambda_{eff}=\lambda- U'\bar{n}(\dfrac{\lambda}{2t})$.

The solutions of  $H^{HF}$ are closely related to the solutions of  
$H$ in the non-interacting case. Yet, the width of the valley states, $\xi$, 
has changed due to the change in $\lambda$.

Moreover, for n.n.-interactions ($U\neq 0$) it is possible to 
use the HF approximation, and obtain the HF 
eigenstates and eigenvalues, which are identical to the non-interacting
solutions, up to the modified 
parameters $\tilde t$, and $\tilde \lambda$ 
\cite{kraus14}. The many-body density and the exchange terms 
are proportional to those obtained 
already for the spinless case \cite{kraus14} up to a 
proportionality constant of $1/2$, 
due to the spin degrees of freedom. Therefore,
$<p_{j,s}>\approx \frac{1}{2} 
\bar{p}(\frac{\lambda}{2t}\cos(2\pi\epsilon j+\phi))$. 
Between the potential valleys this can be approximated by 
$<p_{j,s}>\approx \frac{1}{2} \bar p(\frac{\lambda}{2t})$.

We can now write the HF. Hamiltonian with both Hubbard and n.n. interactions:
\begin{equation}
\begin{aligned}
&H^{HF}=\sum_{s\neq s'=\uparrow,\downarrow}\sum_{j=1}^L \big[t_{eff}(c^{\dagger}_{j,s}c_{j+1,s}+h.c.)\\
&+t'c^{\dagger}_{j,s}c_{j,s'}+\big(\lambda_{eff} \cos(2\pi bj+\phi)+\\
&\frac{1}{2} U+\frac{1}{4}U'\big)n_{j,s}\big],
\end{aligned}
\end{equation}
with $t_{eff}=t+\frac{1}{2}U\bar{p}(\frac{\lambda}{2t})$ and 
$\lambda_{eff}=\lambda+(2U- U')\bar{n}(\dfrac{\lambda}{2t})$.

We again can solve the system with the modified parameters, 
and obtain the HF eigenvalues and eigenstates,
\begin{equation}
\begin{aligned}
E^{HF}_{val}&=\pm \sqrt{8n}\dfrac{t_{eff}}{\xi}\pm_st'+
\frac{1}{2}(U+\frac{1}{2}U')\\
|l_{z,i}>&\approx (\pi\xi^2)^{-\frac{1}{4}}\sum_{j=1}^L (s_z)^j\bold{S}_j 
e^{-\frac{(j-l_z)^2}{2\xi^2}}|j,i>,
\end{aligned}
\end{equation}
where the Gaussian decay parameter $\xi=\xi(\frac{t_{eff}}{\lambda_{eff}})$ 
is modified due to the effective values taken by $\lambda$ and $t$. 
$\xi^2$ is multiplied by a numerical constant equal to $1.16$ 
as in \cite{kraus14}.

Projecting the HF Hamiltonian on the central band yields
\begin{equation}
\begin{aligned}
&H_{central}^{HF}=-\bar t^{HF} \sum^{L_z}_{z=1}
\sum_{s\neq s'=\uparrow,\downarrow}(-1)^zc^{\dagger}_{l_z,s}c_{l_{z+1},s}+h.c.\\
&+t'c^{\dagger}_{l_z,s}c_{l_z,s'}+\frac{1}{4}(2U+U')c^{\dagger}_{l_z,s}c_{l_z,s}.
\end{aligned}
\end{equation}
The eigenvalues and the eigenstates of the central band are then given by:
\begin{equation}
\begin{aligned}
 &E^{central}(k)=(-1)^{n+1}2\bar t^{HF}\cos(k)+ \frac{1}{4}(\pm 4t'+2U+U'),\\
&|k,i>=L_z^{-1/2}\sum^{L_z}_{z=1}S_ze^{ikz}|l_{z,i}>, k=\frac{2\pi n}{L_z}, 
n=1,..,L_z,
\label{eig}
\end{aligned}
\end{equation}
with
$L_z=\lfloor2|\epsilon|L\rfloor$ the number of valley states. 
The hopping amplitude $\bar t^{HF}$ is given by
\begin{equation}
\bar t^{HF}\approx e^{-\frac{1}{4\xi^2\epsilon}}\big(2t_{eff}e^-\frac{1}{4\xi^2}
\sinh\big(\frac{1}{4\xi^2|\epsilon|}\big)-\lambda_{eff}e^{-(\pi\epsilon\xi)^2}\big).
\label{Eq.thf}
\end{equation}
Thus the inverse compressibility $\Delta_2(N)$ can be calculated 
using (\ref{Eq:D2}) and the eigenvalues 
are presented in Eq. (\ref{eig}). 

\begin{figure}[ht!]
\centering
\includegraphics[width=90mm]{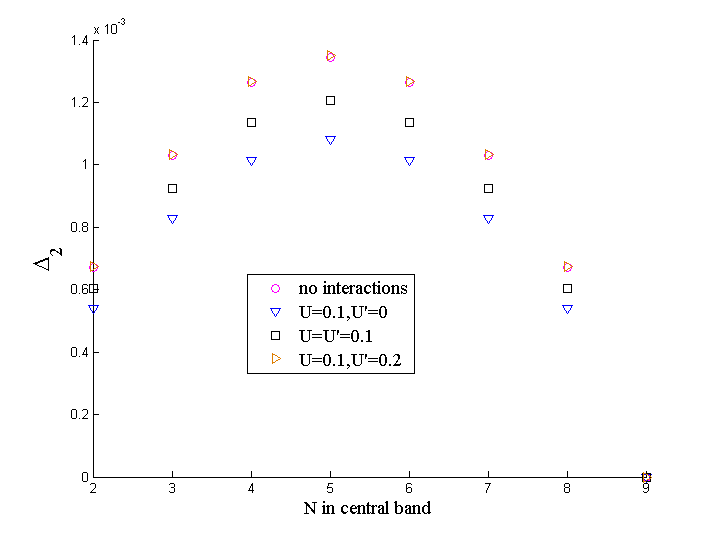}
\caption{The variation of the inverse compressibility $\Delta_2(N)$ of 
the lower central band states with the n.n. interaction ($U$) and the Hubbard 
interaction ($U'$). $\Delta_2(N)$ decreases with $U$, which is in line 
with the results of Ref. \onlinecite{kraus14} and increases with $U'$. 
Thus, the Hubbard interaction delocalizes the particles, smearing their 
wave functions and increasing the amount of energy needed for adding 
another particle to the system.}
\label{delta2vsN}
\end{figure}

As shown in Fig. \ref{delta2vsN}, $\Delta_2(N)$ decreases with the 
n.n.-interaction $U$, in agreement with the case of spinless fermions 
\cite{kraus14}. However, the Hubbard interaction $U'$ enhances $\Delta_2(N)$. 
As was shown in Eq. (\ref{hfhamiltonian}), the Hubbard interaction reduces 
the value of the effective Harper potential amplitude, $\lambda_{eff}$. 
The decrease in $\lambda_{eff}$ increase the width of the Gaussian 
wavefunctions. Thus, there is more overlap between different states and 
therefore any change of configuration in the system, such as adding 
another particle, requires more energy. For $U=2U'$ the system returns to
the non-interacting Hamiltonian value of $\Delta_2(N)$. 
The interplay between $U$ 
and $U'$ determines weather $\Delta_2(N)$  will be larger ($U<2U'$) 
than its non-interacting value or smaller ($U>2U'$) than it.

For an intuitive understanding let us revisit Fig. \ref{spinebands}. 
The states which occupy the lowest energy band reside in the valleys of 
potential. When the Hubbard interaction is turned on, occupying these 
states become too costly in energy for some of the spins. In order to 
avoid the Hubbard interaction they tend to occupy the surroundings 
of potential peaks, where there are less spins to interact with. 
This tendency delocalizes the Gaussian wavefunctions. However, since 
only half of the 
particles participate in the interaction between the opposite spins  
it is less significant (by a factor of $\frac{1}{2}$) than $U$.



An exception to this behavior is found 
for the state at edge of the lower splitted band.
As detailed earlier, due to the spin flipping amplitude $t'$,
a gap of size $2 t'$ opens between the lower central band
occupied by $\frac{1}{\sqrt{2}}(\uparrow+\downarrow)$ states
and the higher central band with states corresponding to 
$\frac{1}{\sqrt{2}}(\uparrow-\downarrow)$ .
$\Delta_2(N)$ decreases with $U'$ and increases with $U$ at the edge,
similar to the behavior observed close to the half-filling point of the 
1D Hubbard model \cite{usuki}. 



\section{Next-nearest neighbors interactions}

In order to understand the behavior of the compressibility for a system 
with long range interactions, 
we consider here the influence of next nearest neighbors interaction. 
For simplicity, we discuss spinless 
fermions. The results of this section can be easily extended for fermions 
with spin using the methods 
described in the previous section.  

The Hamiltonian is given by:
\begin{equation}
\begin{aligned}
&H=\sum_{j=1}^L 
\big[t(c^{\dagger}_{j}c_{j+1}+h.c.)
+\lambda \cos(2\pi bj
+\phi)n_{j}\\
&+Un_{j}n_{j+1}
+U_2n_{j}n_{j+2})\big],
\end{aligned}
\end{equation}
and the mean-field approximation yields
\begin{equation}
\begin{aligned}
&\sum_{j=1}^Ln_{j+2}n_j\approx\sum_{j=1}^L(<n_{j+2}>+<n_{j-2}>)n_j\\
&-<n_j><n_{j+2}>-<\tilde{p_j}>c^{\dagger}_{j+2}c_j\\
&+h.c+|<\tilde{p_j}>|^2,
\end{aligned}
\end{equation}
where $<n_j>$ is the (already known) background density. 
The background exchange energy is $<\tilde{p_j}>\equiv <c_j^{\dagger}c_{j+2}>$.
Here we ignore constant terms, since they do not contribute to $\Delta_2$.
Using the known value of $<n_j>|_{\epsilon=0}$, 
\begin{equation}
\begin{aligned}
&\sum_{j=1}^L(<n_{j+2}>+<n_{j-2}>)=\\
&\sum_j(1-2\bar n(\frac{\lambda}{2t})\cos(2\pi bj+\phi)).
\end{aligned}
\end{equation}
Interestingly, the exchange term disappears 
(the calculation appears in the appendix) resulting in
\begin{equation}
 <\tilde{p_j}>=0.
\end{equation}

This structural robustness can be attributed to the symmetry 
of the non-interacting Hamiltonian's 
wavefunctions used in the calculation. Thus, the additional 
interaction only changes 
the value of $\lambda_{\rm eff}$ without changing the structure 
of the HF Hamiltonian.
The effective Hamiltonian becomes
\begin{equation}
H_{central}^{HF}= \sum^{L_z}_{z=1}-\bar t^{HF}(-1)^zc^{\dagger}_{l_z}c_{l_{z+1}}+h.c.,
\end{equation}
where $\bar{t}^{HF}$ given by Eq. ({\ref {Eq.thf}}) 
with $t_{eff}=t+\frac{1}{2}U\bar{p}(\frac{\lambda}{2t})$ and 
$\lambda_{eff}=\lambda+(2U-2U_2)\bar{n}(\dfrac{\lambda}{2t})$. Here
we ignored on-site terms, which just lead to an over all energy shift.


We also calculate $\Delta_2(N)$ using DMRG~\cite{white92,dmrg}, for 
the following parameters: $b=\sqrt{30}$ (corresponding to $\epsilon 
\approx -0.023$) and $\phi=0.7 \pi$. 
The length of the system is $L=200$, and we calculated the ground state 
energy $\mathcal{E}(N)$ for each number of electrons $N=91,92,\ldots ,108$. 
For $t=1$, the potential amplitude was chosen as $\lambda=0.7$, which 
results in a flat central band, with the typical $\Delta_2$ greater 
than the numerical accuracy. Interaction strengths of 
$U=0$, $U_2=0$ and $U=0.1$ with $U_2=0,0.025,0.05,0.075$ are considered. 
The boundary conditions are open, since it significantly improves 
accuracy~\cite{white92} and we retain $384$ target states. 
The accuracy of $\Delta_2$ is about $\pm 1 \cdot 10^{-4}t$ and the 
discarded weight is $\sim 10^{-7}$.

\begin{figure}[ht!]
\centering
\includegraphics[width=90mm]{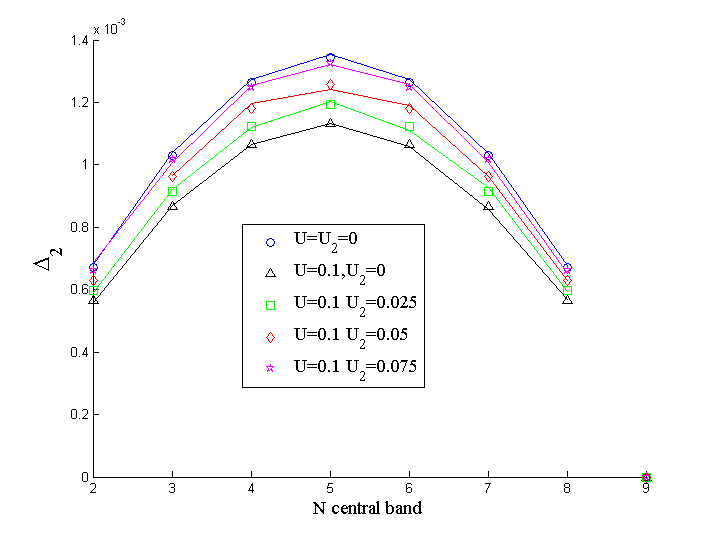}
\caption{The variation of the inverse compressibility $\Delta_2(N)$ 
in the central band of the spinless Harper model 
with the n.n. interaction ($U$) and the next n.n. 
interaction ($U_2$). $\Delta_2(N)$ increases with the next n.n. interaction, 
since the interaction broadens the Gaussian wave functions. 
Thus, adding a particle to the system has a non-local effect, 
and therefore it costs more energy. HF analytic results denoted by symbols, 
DMRG results denoted by straight lines. The DMRG numerical results are 
in agreement to the analytic results we get using the HF method.}
\label{delta2vsN2}
\end{figure}

The resulting change in the compressibility can be viewed in 
Fig. \ref{delta2vsN2}.
Comparing the analytic values to the results obtained using the numerical 
DMRG results, we find good agreement between the two methods. Here the \
Gaussian decay parameter $\xi^2$ is modified according to 
$\xi^2\rightarrow 1.16\xi^2(1-0.4U_2)$. The $1.16$ factor arise 
from using the linear approximation of the potential also between the
valleys, leading to a too-fast decay of the wave function as
was discussed for the n.n interactions \cite{kraus14}.
For the n.n. interaction an additional linear dependence
of $\xi$ on $U_2$ is needed. It seems that 
the longer-range interaction results in an additional correction
of the wave function behavior in the valleys.

With the additional interactions the compressibility ($1/\Delta_2$) 
decreases. Intuitively, the increase in the value of $\lambda$ 
due to the interaction results 
a decrease in the Gaussian decay parameter $\xi^2$, which results in a greater 
overlap between the wavefunctions. This can be interpreted as a change in 
the local nature of the system due to the next n.n.-interactions 
which delocalizes the wavefunctions. Thus, adding another particle 
costs more energy. 
This additional energy cost is reflected in the growth of $\Delta_2(N)$.

\begin{figure}[ht!]
\centering
\includegraphics[width=90mm]{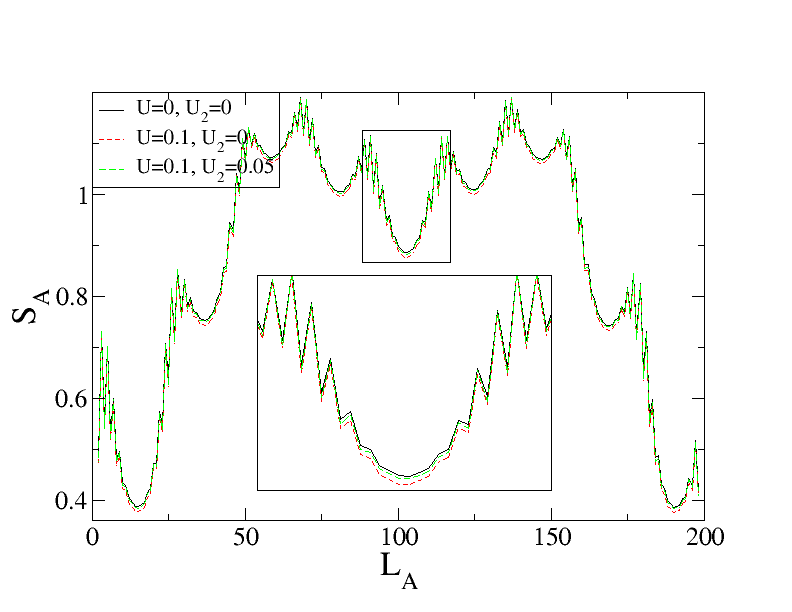}
\caption{The entanglement entropy, $S_A$, as function of the bisection
point $L_A$, for the ground-state with $102$ particles (i.e.,
the $N=7$ state in the central band of the spinless Harper model with
$200$ sites) 
with different n.n. interactions ($U$) and the next n.n. 
interactions ($U_2$). It is apparent that maximums in $S_A$ correspond
to the Gaussian localized states, or the edge states, and that $S_A$ 
around the maximums is not influenced by the interactions. The 
entanglement minimum between
the Gaussian states though are influenced by the interactions.
When $U$ increases (resulting in a decrease in $\Delta_2(N)$) the entanglement
decreases. On the other hand, when $U_2$ increases 
(resulting in an increase in $\Delta_2(N)$) the entanglement is enhanced.
A zoom into the central minimum is presented in the inset.}
\label{eeharp}
\end{figure}

The opposite behavior between the n.n and next n.n. interactions can also
be observed in the behavior of the bipartite entanglement entropy.
The entanglement entropy of a system in a pure state
$|\Psi\rangle$ is defined as the von Neumann entropy of the reduced
density matrix of region A,
$\hat \rho_{A}={\rm Tr}_{B}|\Psi\rangle\langle \Psi |$, where the
degrees of freedom of the rest of the system (region B) are traced out,
resulting in
\begin{eqnarray}
S_{A}=-{\rm Tr} \left( \hat \rho_{A} \ln \hat \rho_{A}\right)
\label{ee}
\end{eqnarray}
For the 1D Harper model the system is divided between regions
A and B, where region A is of length $L_A$ while region B is the remaining
$L-L_A$ sites.

The entanglement entropy for a typical state in the central band is depicted in
Fig. \ref{eeharp}. 
The behavior of $S_A$ is non-monotonous, quite different than the entanglement
entropy of a clean wire, and has several intriguing features.
Here we will concentrate on the feature directly
pertaining to the compressibility. The most obvious feature are the 
peaks appearing in $S_A(L_A)$.
It is apparent that the positions of the peaks not immediately
adjacent to the edges correspond
to the positions of the central band states $|l_{z,i}>$.
These peaks are very robust and do not change when the interaction strength
is changed. On the other hand, the
entanglement of the minimum between
the peaks are influenced by the interactions.
When n.n. interactions ($U$) are introduced the 
entanglement in the minimum regions are suppressed (this is clearly
seen in the enlarge segment in Fig. \ref{eeharp}). 
When next n.n. interactions ($U_2$) are added, the entanglement minimum
remains closer to its non-interacting value. This follows exactly
the pattern exhibited by the inverse compressibility 
($\Delta_2(N)$) is reduced. One can speculate that the entanglement 
is related to the extension of the band state $|l_{z,i}>$ into its
nearest neighbor, and thus the suppression of $\Delta_2(N)$ is related
to the suppression of the entanglement. It is interesting whether
it might be possible to directly relate the compressibility to the
entanglement in a similar 
manner to the relation between fluctuations in the number of particles
and entanglement \cite{song12}. This is left for further study.

\section{Discussion}

In this paper we considered the variation of the inverse compressibility 
$\Delta_2(N)$ with respect to repulsive Hubbard interaction and 
next n.n.-interaction  in the central band of the almost staggered 
fermionic Harper model in the vicinity of half-filling. 
The behavior of the central band states is studied using the HF approximation, 
justified by the flatness of this band and its isolation from the 
other bands. For the next n.n.-interaction we also calculated 
$\Delta_2(N)$ using DMRG. The comparison between the two methods promise 
reliable results.  We found both for the Hubbard interaction and for the 
next n.n. interactions an increase in $\Delta_2(N)$, which corresponds to
a decrease in the compressibility of the system. 
Thus, the increase in the compressibility due to the n.n. interactions
is somewhat suppressed once Hubbard or next n.n. interactions are considered.
It is interesting to note the different role played by the Hubbard
interactions for the clean 1D Hubbard model and the Harper model.
For the clean Hubbard model close to 
to the metal-insulator phase transition at half-filling of 1D systems, 
the Hubbard interaction enhance compressibility \cite{usuki}.
This behavior is also manifested for the Harper model close to the
edge of the lower central band. On the other hand, for the rest of
the central band, the Hubbard 
term effectively reduces the strength of the on-site potential in the 
system ($\lambda_{eff}<\lambda$) and thus the energy gaps 
become smaller, weakening the enhancement of compressibility.
Open questions, such as the classification of interaction terms (which 
terms lead to delocalization and decrease in $\Delta_2(N)$, 
and which localize the wavefunctions and increase $\Delta_2(N)$)
and the full understanding of the non monotonous entanglement entropy,
remain for further study.

%

\begin{acknowledgments}
Financial support from the Israel Science Foundation (Grant 686/10) 
is gratefully acknowledged.
\end{acknowledgments}

\section{Appendix}

For the Hamiltonian with $U=\epsilon=0$, the energy spectrum 
of the central band is
 $E_{k,\pm}=\pm\sqrt{4t^2\cos^2(k)+\lambda^2\cos^2\phi}$.
 The corresponding eigenstates are \[\chi^{\dagger}_{k,\pm}=\sqrt{\dfrac{L}{2}}(c_{ek}^{\dagger},c_{ok}^{\dagger})\left( \begin{array}{ccc}
\chi_{ek,\pm}\\
\chi_{ok,\pm}\\
\end{array} \right)\]
where

\[\left( \begin{array}{ccc}
\chi_{ek,\pm}\\
\chi_{ok,\pm}\\
\end{array} \right)=\dfrac{1}{\sqrt{2E_{k,\pm}(E_{k,\pm}-\lambda \cos\phi)}}\left( \begin{array}{ccc}
2t\cos(k)\\
E_{k,\pm}-\lambda\cos\phi\\
\end{array} \right)\]

The exchange energy is given by
$ <\chi_{k,\pm}|c^{\dagger}_{j+2}c_j|\chi_{k,\pm}>=e^{-2ik}(\chi_{ek,\pm}^2+\chi_{ok,\pm}^2)$.

Normalization yields 

 $\chi_{ek,\pm}^2+\chi_{ok,\pm}^2=1$.

Assuming the lower band is fully occupied,

$<c^{\dagger}_{j+2}c_j>|_{\epsilon=0}=\int^{\pi/2}_{-\pi/2}\dfrac{dk}{\pi}<\chi_{k,-}|c^{\dagger}_{j+2}c_j|\chi_{k,-}>=0$.


\begin{thebibliography}{99}

\bibitem{shechtman84}
D. Shechtman, I. Blech, D. Gratias, and J. W. Cahn, Phys. Rev. Lett. 
{\bf 53}, 1951 (1984).

\bibitem{levine84}
D. Levine, and P. J. Steinhardt, Phys. Rev. Lett. {\bf 53}, 2477 (1984).

\bibitem{Harper:1955} P. G. Harper, Proc. Phys. Soc. London A
{\bf 68}, 874 (1955).

\bibitem{AA} S. Aubry and G. Andr\'{e}, Ann. Isr. Phys. Soc.
{\bf 3}, 133 (1980).

\bibitem{lee85}
P.A. Lee and T.V. Ramakrishnan,
Rev. Mod. Phys. {\bf 57}, 287 (1985).

\bibitem{hiramoto92} For a review see H. Hiramoto and M. Kohmoto,
Int. J. Mod. Phys. B {\bf 6}, 281 (1992).

\bibitem{Svetlana} S. Ya. Jitomirskaya, Annals of Mathematics
{\bf 150}, 1159 (1999).

\bibitem{Roati08} G. Roati, C. D’Errico,  L. Fallani,  M. Fattori,  C. Fort,
M. Zaccanti,  G. Modugno,  M. Modugno and  M. Inguscio, Nature {\bf 453}, 895
(2008).

\bibitem{Yoav09} Y. Lahini, R. Pugatch, F. Pozzi, M. Sorel, R. Morandotti,
N. Davidson, Y. Silberberg, Phys. Rev. Lett. {\bf 103}, 013901 (2009).

\bibitem{Chabe08} J. Chab\'e, G. Lemari\'e, B. Gr\'emaud, D. Delande, P. Szriftgiser, J. C. Garreau, Phys. Rev. Lett. {\bf 101}, 255702 (2008).

\bibitem{Modugno10} G. Modugno, Rep.~Prog.~Phys.~{\bf 73},
102401 (2010).

\bibitem{vidal99}
J. Vidal, D. Mouhanna, and T. Giamarchi, Phys. Rev. Lett. {\bf 83},
3908 (1999); Phys. Rev. B {\bf 65}, 014201 (2001).

\bibitem{schuster02}
C. Schuster, R. A. R\"omer, and M. Schreiber Phys. Rev. B {\bf 65},
115114 (2002) .

\bibitem{iyer13}
S. Iyer, V. Oganesyan, G. Refael, and D. A. Huse
Phys. Rev. B {\bf 87}, 134202 (2013).

\bibitem{kraus12}
Y. E. Kraus, Y. Lahini, Z. Ringel, M. Verbin, and O. Zilberberg,
Phys. Rev. Lett. {\bf 109}, 106402 (2012).

\bibitem{kraus12a}
Y. E. Kraus and O. Zilberberg, Phys. Rev. Lett. {\bf 109}, 116404 (2012).

\bibitem{madsen13}
K. A. Madsen, E. J. Bergholtz, and P. W. Brouwer
Phys. Rev. B {\bf 88}, 125118 (2013).

\bibitem{verbin13}
M. Verbin, O. Zilberberg, Y. E. Kraus, Y. Lahini, and Y. Silberberg,
Phys. Rev. Lett. {\bf 110}, 076403 (2013).

\bibitem{xu13}
Z. Xu, L. Li, and S. Chen, Phys. Rev. Lett. {\bf 110}, 215301 (2013).

\bibitem{ganeshan13}
S. Ganeshan, K. Sun, and S. Das Sarma, Phys. Rev. Lett. {\bf 110},
180403 (2013).

\bibitem{grudst13}
F. Grusdt, M. H\"oning, and M. Fleischhauer, Phys. Rev. Lett. {\bf 110},
260405 (2013).

\bibitem{xu13a}
Z. Xu and S. Chen, Phys. Rev. B {\bf 88}, 045110 (2013).

\bibitem{satija13}
I. I. Satija and G. G. Naumis, Phys. Rev. B {\bf 88}, 054204 (2013).

\bibitem{Bloch:2013}
M. Aidelsburger, M. Atala, M. Lohse, J. T. Barreiro, B. Paredes, and I. Bloch,
Phys. Rev. Lett. {\bf 111}, 185301 (2013).

\bibitem{Ketterle:2013}
H. Miyake, G. A. Siviloglou, C. J. Kennedy, W. C. Burton, and W. Ketterle,
Phys. Rev. Lett. {\bf 111}, 185302 (2013).

\bibitem{sivan96}
U. Sivan, R. Berkovits, Y. Aloni, O. Prus, A. Auerbach, and G. Ben-Yoseph
Phys. Rev. Lett. {\bf 77}, 1123 (1996).

\bibitem{berkovits98}
R. Berkovits, Phys. Rev. Lett. {\bf 81}, 2128 (1998).

\bibitem{alhassid00}
Y. Alhassid, Rev. Mod. Phys., {\bf 72}, 895 (2000).

\bibitem{kurland00}
I. L. Kurland, I. L. Aleiner, and B. L. Altshuler, Phys. Rev. B 62, 14886 (2000).

\bibitem{usuki}
T. Usuki,N. Kawakami, A. Okiji, Phys. Lett. 135A, 476 (1989). 

\bibitem{furukawa}
N. Furukawa, M. Imada, J. Phys. Soc. Jpn. 61, 3331 (1992).

\bibitem{assaad}
F. F. Assaad, M. Imada, Phys. Rev. Lett. 76, 3176 (1996).

\bibitem{kraus14}
Y. E. Kraus, O. Zilberberg and R. Berkovits, Phys. Rev. B {\bf 89}, 161106(R)
(2014). 

\bibitem{white92}
S. R. White, \prl {\bf 69}, 2863 (1992);
\prb {\bf 48}, 10345 (1993).

\bibitem{dmrg}
U. Schollw\"{o}ck, Rev. Mod. Phys. \textbf{77}, 259 (2005);
K. A. Hallberg, Adv. Phys. \textbf{55}, 477 (2006).

\bibitem{berkovits05}
R. Berkovits, F. von Oppen, and Y. Gefen
Phys. Rev. Lett. {\bf 94}, 076802 (2005).

\bibitem{song12}
H. F. Song, S. Rachel, C. Flindt, I. Klich, N. Laflorencie, and K. Le Hur,
Phys. Rev. B {\bf 85}, 035409 (2012).

\end{thebibliography}
\end{document}